\documentclass{article}
\usepackage{epsfig,cite,latexsym}
\title{Coulomb blockade effects in anodised niobium nanostructures}
\author{Torsten Henning, D. B. Haviland, and P. Delsing\\
Department of Physics, G\"oteborgs Universitet\\
och Chalmers Tekniska H\"ogskola AB,\\
S-41296 G\"oteborg, Sweden}
\date{cond-mat/9706302 (1997-06-30)\\
to be published in Supercond. Sci. Technol.}
\begin{document}
\renewcommand{\textfraction}{0.1}
\renewcommand{\topfraction}{0.9}
\maketitle
\begin{abstract}
Niobium thin film wires were fabricated using electron beam lithography
with a four layer liftoff mask system, and subsequently thinned by
anodisation. The resistance along the wire was monitored in situ and
trimmed by controlling the anodisation voltage.
Depending on the room temperature sheet resistance,
samples showed
either superconducting or insulating behaviour at low
temperatures. A Coulomb blockade was observed for samples exceeding
6\,k$\Omega$ per square. 
Samples were also made in a single electron transistor-like geometry
with two weak links made by combined angular
evaporation and anodisation. Their current-voltage characteristics 
could be
modulated by a voltage applied to an overlapping gate.
\end{abstract}
\vspace*{3ex}
\section{Introduction} 
The properties of single electron devices \cite{grabert:92:sctbook}
depend strongly on the electromagnetic environment
\cite{delsing:89:sepprl,devoret:90:envprl,ingold:92:sct}.
Often it is important to bias the single electron device with a constant
current rather than with a constant voltage, over a wide frequency
range. This can in principle be achieved by isolating the device from
the large capacitance of the biasing leads by closely surrounding it
with high ohmic, low capacitance resistors \cite{haviland:91:zfpb}. 
Evaporated thin film alloy resistors \cite{kuzmin:91:cpprl} have the
disadvantage of not allowing a tuning of the resistance in
situ.\par 
We have investigated a process where   niobium wires 
were  transformed to thin film resistors by anodisation
\cite{johansen:57:anox,young:61:anox,alkaine:93:anodi1}. 
The use
of anodisation to tune metal film resistors has been mentioned first in
1959 \cite{westernelectric:60:resipat}.
Anodisation is
frequently used to pattern prefabricated multilayers of niobium and
barrier material for junction applications \cite{kroger:81:anodiapl}. It
has been used to weaken niobium strips into Josephson junctions
\cite{ohta:87:wljj}, and to make variable thickness weak links in
combination with a submicron sized anodisation mask
\cite{goto:81:vtb}. Recently, prefabricated single electron transistors
have been miniaturised by anodisation \cite{nakamura:96:acme}.\par
Ultrathin niobium films show transport properties typical for granular
films \cite{wolf:77:granarr}. In such systems, charging effects are
expected to reduce the conductivity in the low bias region at low
temperatures \cite{gorter:51:filmres}. We have found experimentally that
a Coulomb blockade occurs in samples that have been thinned
sufficiently, and fabricated a device based on anodised niobium weak
links that showed a modulation of current-voltage characteristics with an
applied gate voltage. We refer to these samples as samples with two weak
links in single electron transistor (SET)-like geometry.
\section{Sample fabrication} 
\subsection{Anodisation} 
\begin{figure}\begin{center} 
\epsfig{file=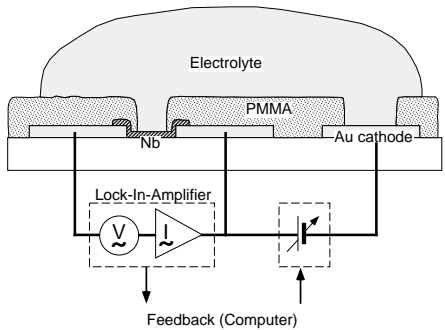,width=0.48\textwidth}
\end{center}\caption{\label{anodisetupfig}%
Setup for Nb microanodisation and resistance trimming (schematic). 
Leads to the resistance monitoring and control electronics were situated
on the chip perimeter, away from the electrolyte droplet.}
\end{figure} 
Figure~\ref{anodisetupfig} shows the experimental setup we used for the
anodisation of microstructures. We spin coated the chip with a
1.8\,$\mu$m thick layer of PMMA and patterned it into an anodisation
mask. Using e-beam lithography for this step provided precision
alignment to the samples. Windows of about ten micrometre width were
opened over the niobium strips that were to be anodised, and over a gold
contact lead electrically isolated from the niobium pattern. This lead
could then be used as cathode for the anodisation. Having such an
integrated cathode near the chip centre reduced the amount of
electrolyte needed to a small droplet of about 1\,mm diameter, 
which minimised
the risk for current leaks and made the handling of the chip
much easier.\par
The design of the anodisation mask was chosen such that it could be used
as a liftoff mask for the deposition of metal forming an overlapping gate
electrode (for the samples in SET-like geometry).\par
As electrolyte, we used a mixture of 30.5\,mmol ammonium pentaborate,
0.87\,mol ethylene glycole and 2.22\,mol water \cite{joynson:67:anodi},
which we found to work well at room temperature although it was originally
intended for use at 120$^\circ$C.\par
To monitor the resistance of the sample during anodisation, we applied
an AC excitation voltage with typical frequency $f=3$\,kHz and
amplitude $V_{\it rms}=4$\,mV and
measured the resulting current with a lock-in amplifier. The voltage had
to be kept small, and especially the  DC component at zero, to avoid
skewing the potential and thus the anodisation profile along the
sample.\par 
\subsection{Resistance trimming} 
The formation of the anodic oxide film, and hence the increase 
of the sample resistance, is irreversible and could be
controlled by regulating the cell voltage. 
Zeroing the voltage holds the
resistance.\par
\begin{figure*}\begin{center} 
\epsfig{file=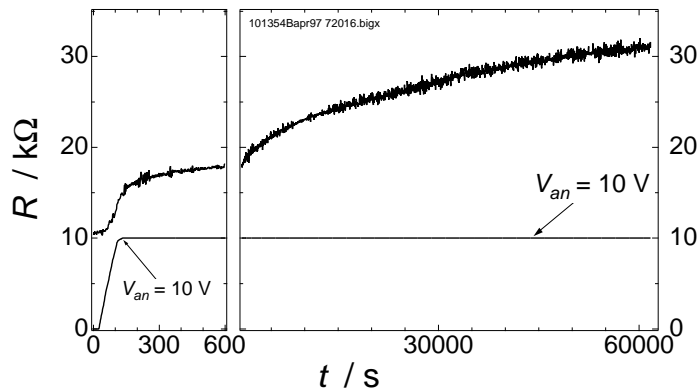,width=0.78\textwidth}
\end{center}\caption{\label{slowanodifig}%
Slow anodisation process. At fixed anodisation voltage, the sample
resistance continued to increase with time.}
\end{figure*} 
Since our anodised areas were very small, we did not attempt to anodise
with a constant current. Instead, we adjusted the cell voltage and its
ramping rate 
based on  the reading of the sample resistance  and its time
derivative. Under such conditions, the oxide thickness is known to
depend not only on the cell voltage, but also on time
\cite{chiou:71:timedep}; an estimation of the oxide film thickness from
an `anodisation constant' of 2.3\,nm/V applicable under constant current
conditions \cite{young:60:anox} 
is incorrect. Depending on the voltage ramping rate, our 20\,nm thick
films were anodised to practically infinite resistance at voltages
between (15\dots 25)\,V.
The
nonlinear behaviour of the oxide thickness is evident from
figure~\ref{slowanodifig}. Here we quickly ramped the anodisation
voltage up to 10\,V and held it constant for more than sixteen
hours. During this time, the resistance increased steadily. This
behaviour can be exploited to achieve a precision in resistance trimming
that is in principle only limited by the accuracy of 
the resistance measurement. We have
tried different anodisation strategies and found that a total process
time of a few hundred seconds gave the most stable and uniform samples,
while providing sufficient trimming precision.\par
Our geometry for the resistor samples were strips about (120\dots
180)\,nm wide and 10\,$\mu$m long, either individual or grouped as
120\,$\mu$m long wire equipped with equidistantly spaced leads allowing
four probe measurements on each segment. The initially 20\,nm thick Nb
films had a sheet resistance of (50\dots 120)\,$\Omega/\Box$, which by
anodisation could be raised to several tens of kiloohms per
square. Strips anodised simultaneously showed the same ratio between
inital and final resistance within ten percent accuracy, up to sheet
resistances of 20\,k$\Omega/\Box$. For higher resistivities, the
uniformity degraded, and at values of 40\,k$\Omega/\Box$, similar strips
could differ in resistance by a factor of 4 or more.\par
\subsection{Patterning} 
The patterning of niobium with a lift-off process requires masks that
can withstand the heat
during
the niobium deposition. Metal stencils supported by a sturdy polymer
\cite{jain:85:nbpbieee} have been reported as a suitable replacement for
conventional all-polymer masks. We employed 
here a modification of a four
layer resist \cite{harada:94:nbset} which 
had been used for the fabrication
of Nb based single electron transistors. It consists of a PMMA top layer
(50\,nm), a germanium mask (20\,nm) supported by hard baked photoresist
(250\,nm), and a PMMA bottom layer (50\,nm) allowing liftoff. The top
layer was patterned by electron beam lithography at 50\,kV acceleration
voltage and 20\,pA beam current, and developed in a mixture of 400\,mmol
isopropanole and 170\,mmol water. The pattern was then transferred to
the Ge mask by rective ion etching (RIE) with carbon tetrafluoride
CF$_4$ as process gas. Subsequent RIE with oxygen through the
photoresist and bottom PMMA layers gave the undercut profile and
suspended bridges necessary for the angular evaporation
technique. Etching times 
were estimated with some margin since we had no
etch end detection.\par
The niobium was deposited in a multipurpose high vacuum system
with an electron gun at a rate of (0.6\dots
0.8)\,nm/s. The sample was neither cooled nor  heated
during evaporation, except by the evaporation itself. Due to the poor
background pressure of $(3\pm 1)\cdot10^{-5}$\,Pa, the niobium had a
residual resistance ratio of only 1.12 between room temperature and
4.2\,K, and a superconducting
transition temperature of the order of 1.5\,K. With the same 
evaporation system and using a similar resist,
Nb based single electron transistors had been fabricated earlier
\cite{harada:94:nbset}.\par 
After the anodisation was completed, the samples were rinsed with
deionised water. For the samples with two weak links in 
a SET-like geometry, a
gate electrode of 50\,nm gold was evaporated, and the mask removed
prior to the low temperature measurements.\par 
\subsection{Angular evaporation} 
\begin{figure}\begin{center} 
\epsfig{file=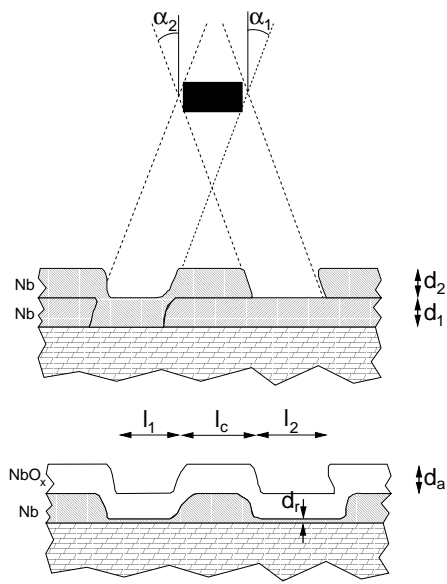,width=0.48\textwidth}
\end{center}\caption{\label{evapfig}%
Shadow evaporation technique for the fabrication of SET-like
samples. Anodisation creates an oxide layer of thickness $d_a$, thinning
the weak links to a thickness $d_r$. Neglected in this sketch are the
deposition of material on the bridge, and the swelling of the film
during anodisation.}
\end{figure}
\begin{figure*}\begin{center} 
\epsfig{file=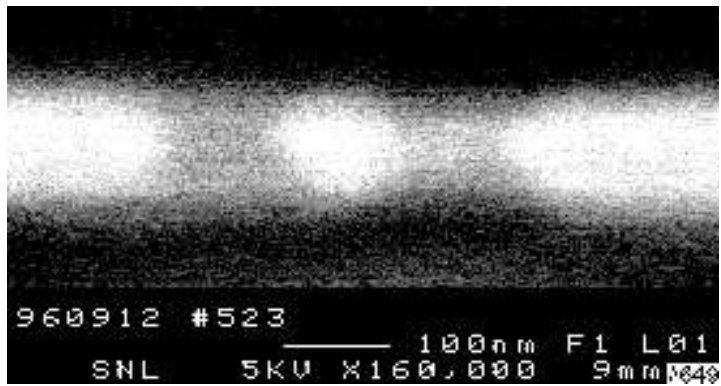,width=0.78\textwidth}
\end{center}\caption{\label{semfig}%
Scanning electron micrograph of a SET-like geometry sample, made by
shadow evaporation, before anodisation.}
\end{figure*}

 
The shadow evaporation technique we used for the fabrication of samples
with two weak links in SET geometry is shown schematically in
figure~\ref{evapfig}. We evaporated two times 20\,nm of Nb, tilting the
sample $\pm 22^\circ$ to the substrate normal. Figure~\ref{semfig} is a
scanning electron micrograph of a structure produced by this method,
before anodisation.
Typical dimensions for these samples were line widths of (90\dots
120)\,nm, length of the weak links $l_{1,2}=(60\dots 100)$\,nm and a
length of the island separating these weak links $l_c=(40\dots
120)$\,nm. This method is capable of producing smaller islands than
could be defined lithographically with similar processing. On the other
hand, the symmetry of the two weak links depends on the equality in
thickness $d_1=d_2$ of the two metal layers deposited in sequential
evaporations. The film thicknesses were monitored with a water cooled
crystal monitor.
A reproducibility in thickness of about 10\,\%
was estimated from control
evaporations under similar conditions and stylus-method profilometry of
the control samples.\par
\section{Results} 
\subsection{Resistor samples} 
Transport measurements were carried out in a dilution refrigerator at
base temperature well below 50\,mK. Details about the cryostat regarding
the filtering of leads have been published elsewhere
\cite{haviland:96:jvst}. The bias voltage was applied symmetrically with
respect to ground and fed to the sample via high ohmic bias resistors on
top of the cryostat. The voltages over these resistors and over the
sample were picked up by low noise amplifiers \cite{delsing:90:thesis}
and registered with digital voltmeters.\par
\begin{figure*}\begin{center} 
\epsfig{file=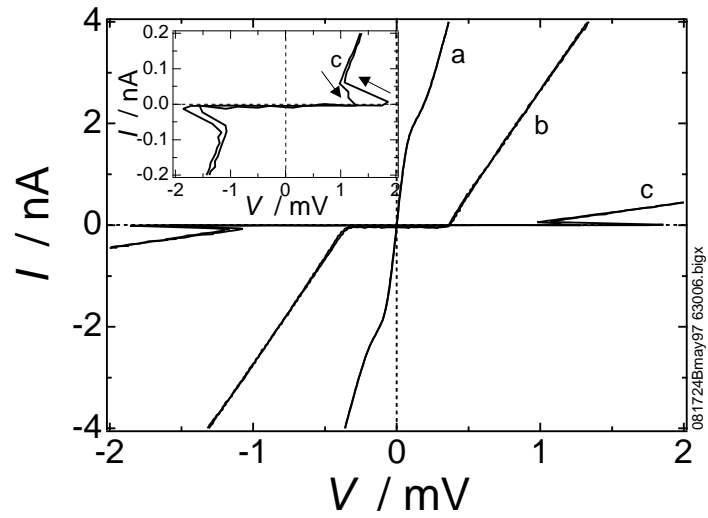,width=0.78\textwidth}
\end{center}\caption{\label{ivcfig}%
Current-voltage characteristics (IVC) of resistor samples. With
increasing sheet resistance at high bias or high temperature, the IVC
changed from a supercurrent remnant (a: 1.5\,k$\Omega/\Box$) to a sharp
Coulomb blockade (b: 8\,k$\Omega/\Box$). Samples with very high
resistance showed a backbending IVC (c: 40\,k$\Omega/\Box$).
Sample dimensions $10\times 0.15\,\mu$m$^2$, $T<50$\,mK, 
$B_{\it ext}=0$.}
\end{figure*}
\begin{figure}\begin{center} 
\epsfig{file=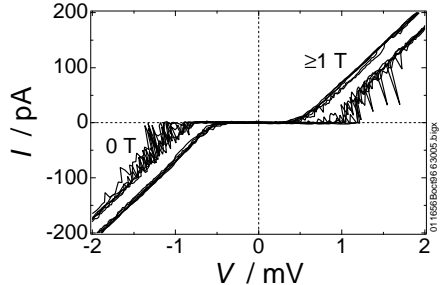,width=0.48\textwidth}
\end{center}\caption{\label{magfig}%
Effect of magnetic field on IVC of high resistive strip sample. The
offset and threshold voltages were reduced and the switching along the
load line was smoothened out as superconductivity was quenched. Several
traces are superimposed for each case, $T<50$\,mK.}
\end{figure}

 
Figure~\ref{ivcfig} summarises the kinds of current-voltage
characteristics (IVC) we observed in our resistor wire samples. Low
resistive strips showed a remnant of a supercurrent (trace a). For
higher (sheet) resistance, we observed a Coulomb blockade (trace
b). Very high resistive samples showed not only a sharp threshold
voltage, but a backbending IVC (trace c). Such a backbending is a
characteristic of the Coulomb blockade of Cooper pair tunnelling
\cite{geerligs:89:jjaprl} in arrays of Josephson junction arrays
\cite{chen:92:ps}. The backbending alone could be attributed to heating
effects, but we also regularly observed  a vivid switching of the IVC
along the bias load line (figure~\ref{magfig}, 
traces labelled `0\,T') that is
typical in arrays of Josephson junctions
\cite{haviland:96:jvst}. 
This switching is random, but occurs between reproducible envelopes in
subsequent sweeps.
Further evidence for an influence of
superconductivity on the IVC near the Coulomb blockade threshold
threshold is the fact that a
magnetic field suppressing superconductivity also suppresses the
switching, as we see in figure~\ref{magfig}.\par
To quantify the Coulomb blockade, we considered the offset voltage. Since
the current-voltage characteristics were nonlinear over many decades in
voltage, it was not enough to simply extrapolate the tangent to the IVC
at the edges of the sweep range. Instead, we computed the local offset
voltage $V_{\it off}(V)$ 
\cite{wahlgren:95:prb}
as the voltage value at the intersection of the
tangent to the IVC in $V$ with the voltage axis,
\begin{equation}
V_{\it off}(V)=V-I(V)\;
\left.\frac{dV^\prime}{dI}\right|_{V}.
\end{equation}
Around zero bias, $V_{\it off}(V)$ was approximately a linear function of
$V$.
The extrapolation of $V_{\it off}(V)$ to zero bias, $V_{\it off}^0$, is
a measure for the strength of the Coulomb blockade. For single
ultrasmall junctions, this gives the Coulomb blockade in the global
rule, where the whole electromagnetic environment influences the
blockade.\par
\begin{figure*}\begin{center} 
\epsfig{file=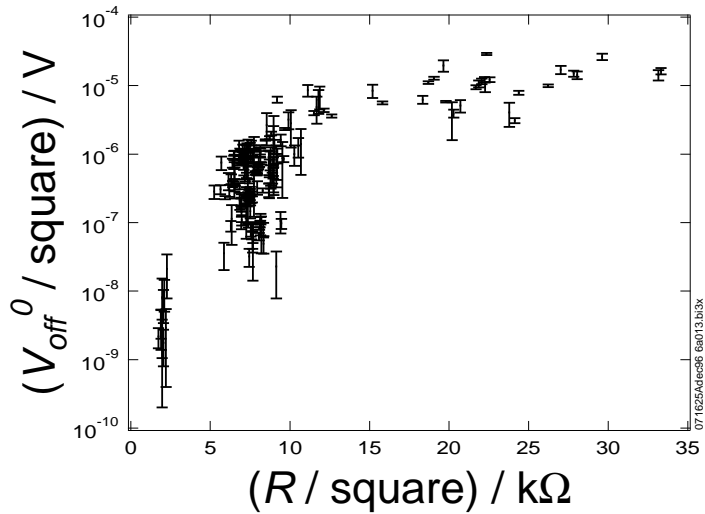,width=0.78\textwidth}
\end{center}\caption{\label{onsetfig}%
Onset of Coulomb blockade (CB) in resistor samples.
Both $V_{\it off}^0$ and the resistance $R$ are normalised
to the number of squares. 
The CB is suppressed below 6\,k$\Omega/\Box$.
Errors were estimated from four extrapolations per 
(bidirectionally swept) IVC.}
\end{figure*} 
For 187 strips with lengths between ten and 120\,$\mu$m, we have
measured the IVC without applied external magnetic field. 
The relation
between $V_{\it off}^0$ and the sample resistance is 
shown in figure~\ref{onsetfig}, where both
quantities are normalised to the number of squares in the film and
plotted against each other. The Coulomb
blockade set in around a sheet resistance 
per square  compatible with the
quantum resistance 
\begin{equation}
R_Q=R_K/4=h/(4e^2)\approx 6.4\,\mbox{k}\Omega.
\end{equation}
This supports the 
notion that there should be a universal sheet
resistance $R_Q/\Box$ for the superconductor-insulator transition in thin
metallic films \cite{belitz:94:rmp,liu:93:sitprb}.
\subsection{Samples in SET-like geometry} 
Although our resistor wire samples showed a Coulomb blockade, we did not
observe a modulation of the conductance with an applied gate voltage, as
reported e.\,g. for high resistance In$_2$O$_{3-x}$ wires
\cite{chandrasekhar:94:jltp}. We assume that the analogy 
between our wire samples and an array of
junctions holds, and that the number of islands involved was so large
that such effects averaged out below the measurement noise level.\par
Therefore, we fabricated the samples with two weak links and an overlapping
gate described above, whose geometry resembles that of a single electron
transistor (SET), where the oxide barriers have been replaced by short
strips of niobium thinned by anodisation.\par
\begin{figure}[t]\begin{center} 
\epsfig{file=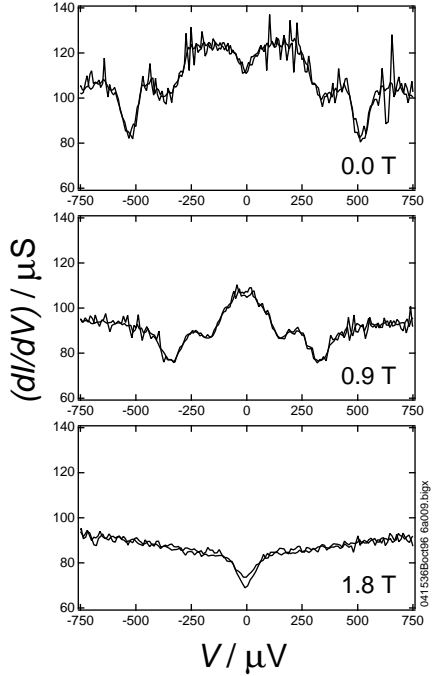,width=0.48\textwidth}
\end{center}\caption{\label{dipsfig}%
Coulomb blockade in a SET-like sample. As superconductivity was quenched
by an external field, the off-zero-bias conductance peaks disappeared, and
a Coulomb blockade for 
single electrons remained. $T\approx40$\,mK.}
\end{figure} 
Samples that had a complicated current-voltage
characteristic  were susceptible to a modulation of this
characteristic by a gate voltage. Figure~\ref{dipsfig} is a plot of the
differential conductivity of such a sample as a function of bias
voltage. The  off-zero bias conductance peaks and dips
moved towards zero bias with an increasing magnetic field and converged
into a single conductance dip at an external field of 1.4\,T. The latter
zero bias conductivity dip is obviously indicating the Coulomb blockade
of single electron tunnelling, while the off-zero bias structures are
presumably
caused by the superconducting gap. The reduction of the 
differential conductivity
at zero bias in the field free case (top panel in figure~\ref{dipsfig})
may be attributed to the Coulomb blockade of  Cooper pairs
\cite{haviland:91:zfpb}.\par
On the other hand,
samples that showed a  Coulomb blockade with a very sharp threshold had 
current-voltage characteristics
that could not be modulated with a gate voltage. 
Samples that had a
very weak Coulomb blockade, resulting in a simple dip in the
differential conductivity versus bias voltage characteristic
around zero bias without further structure,
were insensitive to a gate voltage
as well. Scanning microscope inspection of
these samples often suggested that the asymmetry between the two weak
links was non-negligible.\par
\begin{figure}[t]\begin{center} 
\epsfig{file=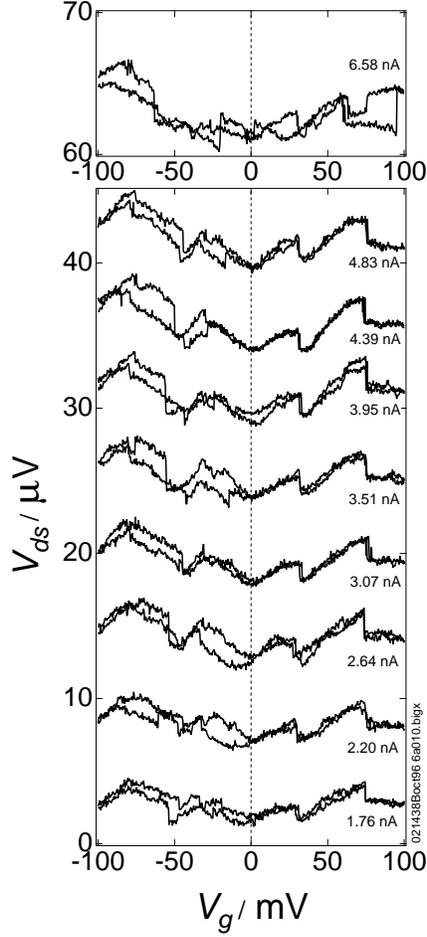,width=0.48\textwidth}
\end{center}\caption{\label{sawfig}%
Control curves for a sample with SET-like geometry. The drain-source
voltage $V_{\it ds}$ at different current bias points oscillated as the
voltage $V_g$ applied to a top gate varied. $T\approx30$\,mK, $B_{\it
ext}=2$\,T.} 
\end{figure}

To measure the response of the sample
from figure~\ref{dipsfig} to the gate voltage, we quenched
superconductivity by applying a field of 2\,T, and biased the sample via
high ohmic resistors at a series of practically constant drain-source
currents. We then measured the drain-source voltage drop $V_{\it ds}$
while sweeping the gate voltage $V_g$ up and down. Figure~\ref{sawfig}
gives 
the resulting control curves ($V_{\it ds}$ vs. $V_g$ 
at constant $I_{\it ds}$).\par
The  influence of the gate voltage is obvious for low drain-source
currents (bottom panel). At other points along the $I_{\it ds}$-$V_{\it
ds}$ characteristic, namely for oppositely directed current, or higher
current (top panel), the correlation between the control curves
resulting from upward and downward sweep in gate voltage was less
pronounced. \par
The large period of the observed sawtooth oscillations would correspond
to a total island capacitance of only 3\,aF, if one assumed that the
sample acted as a single electron transistor. This capacitance estimate
is unrealistically low. Control curves with such a large periodicity in
gate voltage are, however, typical for systems of multiple tunnel
junctions, e.\,g. those produced by the step-edge cutoff technique
\cite{altmeyer:95:apl}, or nanofabricated silicon wires
\cite{smith:97:siwirepreprint}. \par
\begin{figure*}\begin{center} 
\epsfig{file=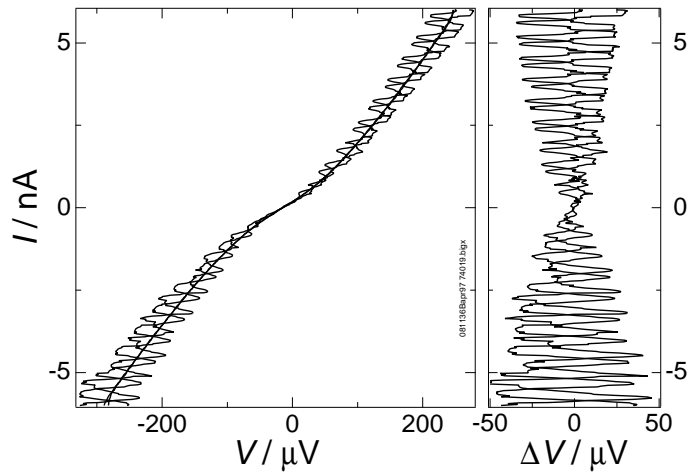,width=0.78\textwidth}
\end{center}\caption{\label{modufig}%
IVC of a SET-like sample with gate modulation. The amplitude 
(rms) of the gate
voltage was 120\,mV and its frequency 40\,times the frequency of the
bias voltage sweep. The right panel gives the deviation of $V$ from the
unmodulated IVC. $T\approx 50$\,mK, no external magnetic field.}
\end{figure*} 
In another sample, we managed to modulate the drain-source
current-voltage characteristic  by applying a gate
voltage. For the measurement shown in figure~\ref{modufig}, we ramped the
bias current up and down
over one period with a frequency of 8\,mHz. Simultaneously, a sine
shaped gate voltage with an amplitude of 120\,mV and a frequency of
322\,mHz was applied. At this amplitude, the whole modulation range of
the IVC was covered. The left panel of figure~\ref{modufig} shows the
modulated and the
unmodulated IVC for comparison, and the difference in drain-source
voltage between unmodulated and modulated IVC is plotted in the right
panel. Unfortunately, this sample was destroyed before we
could map out the modulation range to higher drain-source voltages. 
At 50\,mK and a gate voltage amplitude of 200\,mV, the measured
modulation of the IVC translates into an
average transistor gain of about 1/700 at the edges of the
measured bias region.
\par
The voltage swing $\Delta{}V$ dropped to about one fourth when the
temperature was raised from 50\,mK to 1.1\,K.\par
The insulation resistance between drain-source and gate was at least
30\,G$\Omega$, our measurement limit, for gate voltages up to
$V_g\approx 1.5$\,V. There, a measurable current flow through the gate
set in, reaching about 5\,nA at $V_g=3$\,V.
\section{Conclusion} 
We have demonstrated that anodic oxidation of nanofabricated niobium
thin film wires can be used to produce resistors of several hundred
kiloohms on a length of ten micrometres. This technique is intrinsically
limited by the onset of a Coulomb blockade when the sheet resistance
per square
exceeds the quantum resistance 6\,k$\Omega/\Box$. The anodised wires show
transport properties typical of an array of ultrasmall Josephson
junctions, where superconducting effects coexist with charging
effects. The low temperature 
current-voltage characteristics
of these samples show a
superconductor-insulator transition where the degree of anodisation is
the tuning parameter.\par
Placing short anodised areas with the aid of lithographic techniques,
one can fabricate transistor-like samples whose 
current-voltage characteristics can be modulated by
a gate voltage. More efforts are needed to further characterise the
complex niobium-niobium oxides system created by anodisation and
comprising the resistors and single electron transistor-like
structures.\par
\section*{Acknowledgements} 
Samples were fabricated in the Swedish Nanometer Laboratory,
G\"oteborg. T.\,H. gratefully acknowledges financial support by the
German Academic Exchange Service trough HSP II. This research is part of
ESPRIT project 9005 SETTRON. We were supported by Swedish NFR, TFR,
the Wallenberg Foundation, and the Commission of the
European Communities. 

\end{document}